\documentclass{article}
\usepackage{graphicx}
\usepackage[centertags]{amsmath}
\usepackage{amsfonts}
\usepackage{amssymb}
\usepackage{amsthm}

\title{\Large Parametric representation of wave propagation\\ 
in nonuniform media (both in transmission\\ and stop bands)}

\author{A. Popov$^{\sharp}$ and V. Kovalchuk$^{\flat}$\\
\\
${}^{\sharp}$ Pushkov Institute of Terrestrial Magnetism, Ionosphere\\
and Radio Wave Propagation, Russian Academy of Sciences,\\
Troitsk, Moscow region, 142190 Russia\\
\\
${}^{\flat}$ Institute of Fundamental Technological Research,\\
Polish Academy of Sciences,\\
$5^{\rm B}$, Pawi\'{n}skiego str., 02-106 Warsaw, Poland\\
\\
{\it e-mails: popov@izmiran.ru, vkoval@ippt.gov.pl}}

\begin{document}

\maketitle

\begin{abstract}
An analytical approach based on the parametric representation of the wave propagation in nonuniform media was considered. In addition to the previously developed theory of parametric antiresonance describing the field attenuation in stop bands, in the present paper the behaviour of the Bloch wave in a transmission band was investigated. A wide class of exact solutions was found and the correspondence to the quasi-periodic Floquet solutions was shown.

\end{abstract}

\section*{Introduction}

In this paper we concentrate on the analytical theory of linear propagation in nonuniform media based. The difficulty of this problem comes out even in the one-dimensional (1D) case, as the ordinary differential equation (ODE)
\begin{equation}\label{eq.1}
w^{\prime\prime}(x)+q^{2}(x)w(x)=0, \qquad q(x)=\frac{\omega}{c}n(x),
\end{equation}
has no explicit solution except for some rare model functions $q(x)$. The above ODE describes 1D harmonic waves $\sim \exp\left(-i\omega t\right)$ propagating in a nonuniform dielectric medium with gradually varying refraction index $n(x)$; $c$ is the speed of light in vacuum and $\prime$ denotes the differentiation with respect to $x$. 

Real-valued solutions of ODE (\ref{eq.1}) correspond to standing waves carrying no energy while complex-valued ones describe travelling waves. For $q={\rm const}$ both types are ``sinusoidal" (proportional to $\sin/\cos qx$ or $\exp\left(\pm iqx\right)$). Evidently, variations of the parameter $q(x)$ may change their amplitude and wave form. Moreover, for a periodic modulation of the refraction index $n(x)$ the parametric resonance/antiresonance (i.e., the existence of exponentially increasing/decreasing solutions) may occur.

According to the Floquet theorem \cite{3,4}, for any general periodic modulation $q\left(x+\chi\right)=q(x)$ (or equivalently $n\left(x+\chi\right)=n(x)$), where $\chi$ is the period, there exists the following solution of the equation (\ref{eq.1}) (Bloch wave):
\begin{equation}\label{eq.1.1}
w(x)=\widetilde{w}(x)\exp\left(\pm \mu x\right),
\end{equation}
where $\widetilde{w}(x)$ is a periodic function and the characteristic exponent $\mu$ can be either $(i)$ real or $(ii)$ purely imaginary \cite{4}. The former case (with the minus sign) corresponds to a parametric antiresonance (wave attenuation) in the stop bands (``forbidden zones") of the periodic structure (multilayer interference mirrors, Bragg optical waveguides); in other words it provides localization of the propagating mode by the periodic dielectric cladding. The latter case describes a periodic modulation of the travelling waves.

In many applications the inverse problem (a kind of optimal control) arises when one needs to find such a dependence $n(x)$ that provides the maximal value of the characteristic exponent $\mu$. In our previous papers \cite{1,2,1a} a simple analytical theory of parametric (anti)resonance was developed and explicit formulae for field attenuation in stop bands was derived. However, for the sake of completeness, we need also to know the behaviour of the Bloch wave in a transmission band where the characteristic exponent is purely imaginary. The first attempt to apply the method of phase parameter to the transmission bands has been reported in \cite{2a}.

In the present paper, being a rectification of \cite{2a}, we develop a general approach based on the parametric representation of the wave propagation in nonuniform media (both in transmission and stop bands). A physically correct definition of phase is put forward for a general case of linear wave propagation in a nonuniform medium. A wide class of exact solutions to the 1D wave equation (\ref{eq.1}) is found and the correspondence to the quasi-periodic Floquet solutions is shown.

\section{Phase parameter for the real case}

Let us recall the results obtained in the previous papers of the first author (AP) concerning the parametric representation of the fundamental system of solutions to the wave equation (\ref{eq.1}). If we introduce an admittance function
\begin{equation}\label{eq.1a}
y(x)=\frac{w^{\prime}(x)}{q(x)w(x)},
\end{equation}
i.e., the solution of the equation (\ref{eq.1}) can be represented as follows:
\begin{equation}\label{eq.1.a.1}
w(x)=w_{0}\exp\left[\int y(x)q(x)dx\right],
\end{equation}
where $w_{0}$ is an integration constant, then it is easy to observe that the primary wave equation (\ref{eq.1}) can be equivalently rewritten in the following form:
\begin{equation}\label{eq.1b}
q(x)y^{\prime}(x)+q^{\prime}(x)y(x)+q^{2}(x)\left[1+y^{2}(x)\right]=0,
\end{equation}
i.e.,
\begin{equation}\label{eq.1c}
\int\frac{y^{\prime}(x)dx}{q(x)\left[1+y^{2}(x)\right]}+
\int\frac{y(x)q^{\prime}(x)dx}{q^{2}(x)\left[1+y^{2}(x)\right]}=x_{0}-x,
\end{equation}
where $x_{0}$ is an integration constant. 

If $q(x)\equiv q_{0}$ is a constant, then the above equation (\ref{eq.1c}) reads
\begin{equation}\label{eq.1_1}
\frac{1}{q_{0}}\int\frac{dy}{1+y^{2}}=x_{0}-x.
\end{equation}
We see that the integral in (\ref{eq.1_1}) can be easily integrated with the substitution
\begin{equation}\label{eq.1_1.1}
y={\rm ctg}\: \psi,\qquad dy=-\frac{d\psi}{\sin^{2}\psi},\qquad 1+y^{2}=\frac{1}{\sin^{2}\psi},
\end{equation}
which gives us the solution
\begin{equation}\label{eq.1_6}
\psi={\rm ctg}^{-1}y=\psi_{0}+q_{0}\left(x-x_{0}\right)=
q_{0}\left(x-\widetilde{x}_{0}\right),
\end{equation}
where $\psi_{0}$ is an integration constant and $\widetilde{x}_{0}=x_{0}-\psi_{0}/q_{0}$. 

Therefore, the solution (\ref{eq.1.a.1}) can be rewritten in the equivalent form: 
\begin{eqnarray}\label{eq.1_2}
w(x)&=&w_{0}\exp\left\{q_{0}\int{\rm ctg}\left[ q_{0}\left(x-\widetilde{x}_{0}\right)\right]dx\right\}
\nonumber\\
&=&w_{0}\exp\left\{\int\frac{d\sin\left[q_{0}\left(x-\widetilde{x}_{0}
\right)\right]}{\sin\left[q_{0}\left(x-\widetilde{x}_{0}\right)
\right]}\right\}=\widetilde{w}_{0}\sin\left[q_{0}\left(x-\widetilde{x}_{0}
\right)\right],
\end{eqnarray}
where $\widetilde{w}_{0}$ is new integration constant.

More generally, for an arbitrary function $q(x)$ we can, by analogy to the above reasoning, define a new variable $\psi(x)$ (phase parameter) as follows:
\begin{equation}\label{eq.2}
y(x)={\rm ctg}\: \psi(x).
\end{equation}
Then the equation (\ref{eq.1b}) reads
\begin{equation}\label{eq.2c}
-\frac{q(x)\psi^{\prime}(x)}{\sin^{2}\psi(x)}+q^{\prime}(x)\: {\rm ctg}\: \psi(x)+\frac{q^{2}(x)}{\sin^{2}\psi(x)}=0,
\end{equation}
i.e.,
\begin{equation}\label{eq.2d}
\psi^{\prime}(x)-\frac{q^{\prime}(x)}{2q(x)}\sin 2\psi(x)=q(x).
\end{equation}

If there exists the inversion $x=X(\psi)$, then we can write $w(x)$, $y(x)$, and $q(x)$ as functions of $\psi$, i.e., 
\begin{equation}\label{eq.2b}
w\left[X(\psi)\right]=W(\psi),\quad y\left[X(\psi)\right]=Y(\psi)\equiv {\rm ctg}\: \psi,\quad q\left[X(\psi)\right]=Q(\psi). 
\end{equation}
Then the equations (\ref{eq.1.a.1}) and (\ref{eq.1c}) can be rewritten as follows:
\begin{eqnarray}
W(\psi)&=&w_{0}\sin\psi\exp\left[-
\int\frac{\dot{Q}(\psi)}{Q(\psi)}\cos^{2}\psi\: d\psi\right],\label{eq.1d}\\
X(\psi)&=&x_{0}+\int\frac{d\psi}{Q(\psi)}
-\frac{1}{2}\int\frac{\dot{Q}(\psi)}{Q^{2}(\psi)}\sin 2\psi\: d\psi.
\label{eq.1e}
\end{eqnarray}
Here and below dots denote the differentiation with respect to $\psi$. For any arbitrary chosen function $Q(\psi)$ we obtain explicit expressions for $W(\psi)$ and $X(\psi)$, therefore, for $w(x)=W\left[\psi(x)\right]$. 

If we make a substitution $Q(\psi)=q_{0}\exp G(\psi)$, where $q_{0}$ is some constant, then the above equations (\ref{eq.1d}) and (\ref{eq.1e}) will take the following form:
\begin{eqnarray}
W(\psi)&=&w_{0}\sin\psi\exp\left[-
\int \dot{G}(\psi)\cos^{2}\psi\: d\psi\right],\label{eq.3b}\\
X(\psi)&=&x_{0}+\frac{1}{q_{0}}\left[\int\frac{d\psi}{\exp G(\psi)}
-\frac{1}{2}\int\frac{\dot{G}(\psi)\sin 2\psi\: d\psi}{
\exp G(\psi)}\right].\label{eq.3c}
\end{eqnarray}

In such a way we construct a closed-form parametric solution which reveals some quantitative relations inherent in wave propagation in nonuniform media, not obvious in classical solutions of the corresponding differential equations \cite{3}. In particular, for any periodic refraction index $n(x)=n(x+\chi)$ defined implicitly by a Fourier series, i.e.,
\begin{equation}\label{eq.4}
G(\psi)=a_{0}+\sum^\infty_{m=1}\left(a_{2m}\cos 2m\psi+b_{2m}\sin 2m\psi\right),
\end{equation}
where $a_{0}$, $a_{2m}$, $b_{2m}$ are some constants, we obtain a Floquet solution (\ref{eq.1.1}) with the minus sign in the exponent and the characteristic exponent $\mu=\nu/\chi$ with the explicit formulae for the period $\lambda$ and the attenuation per period $\nu$ \cite{1,2,1a,2a}:
\begin{equation}\label{eq.6}
\chi=\frac{2}{q_{0}}\int^{\pi}_{0}\exp\left[-G(\psi)\right]
\sin^{2}\psi d\psi,\qquad \nu=\int^{\pi}_{0}G(\psi)\sin 2\psi d\psi.
\end{equation}
These analytical relations, giving the very simple description of the wave field attenuation in a periodic structure, are useful for the optimal design of multilayer mirrors and Bragg fibre claddings \cite{1}. However, from the theoretical point of view this solution remains incomplete until a similar parametric representation is found for the propagating waves in the corresponding transmission bands of a periodic medium.

\section{Complex case}

For a complex wave also is possible to define a phase parameter $\psi(x)$, which obviously must be a homogeneous function of $w(x)$ and $w^{\prime}(x)$. 

Let us observe that using the identity
\begin{equation}\label{eq.8b}
\frac{y(x)+i}{y(x)-i}=\frac{{\rm ctg}\: \psi+i}{{\rm ctg}\: \psi-i}=
\frac{\cos\psi+i\sin\psi}{\cos\psi-i\sin\psi}=\exp\left(2i\psi\right)
\end{equation}
Eq.\ (\ref{eq.2}) can be equivalently rewritten as follows:
\begin{equation}\label{eq.8}
\psi(x)={\rm ctg}^{-1}y(x)=\frac{1}{2i}\ln\frac{y(x)+i}{y(x)-i}=
\frac{1}{2i}\ln\frac{w^{\prime}(x)+iq(x)w(x)}{w^{\prime}(x)-iq(x)w(x)}.
\end{equation}

Then let us define the quasi-phase parameter $\psi(x)$ of a complex wave function $w(x)$ as follows:
\begin{equation}\label{eq.9}
\psi(x)=\frac{1}{2i}\ln\frac{w^{\prime}(x)+iq(x)w(x)}
{w^{\ast\prime}(x)-iq(x)w^{\ast}(x)}=
\int\left\{q(x)+\frac{q^{\prime}(x)}{q(x)}\frac{{\rm Re}\left[y(x)\right]}
{\left|y(x)+i\right|^{2}}\right\}dx.
\end{equation}
The second part of Eq.\ (\ref{eq.9}) can be obtained from the first one by the direct calculation of the integral representation of the logarithm, i.e.,
\begin{equation}\label{eq.12a}
{\rm Re}\left[\frac{1}{i}\int\frac{d(w^{\prime}+iqw)}{w^{\prime}+iqw}\right]
={\rm Re}\left[\frac{1}{i}\int\frac{w^{\prime\prime}+
i(qw^{\prime}+q^{\prime}w)}{w^{\prime}+iqw}dx\right]
\end{equation}
and now, using Eq.\ (\ref{eq.1}) and the facts that ${\rm Re}\left(iz\right)=-{\rm Im}(z)$, ${\rm Im}\left(iz\right)={\rm Re}(z)$, we finally obtain that Eq.\ (\ref{eq.12a}) can be rewritten as follows:
\begin{equation}\label{eq.12b}
{\rm Im}\left[\int\frac{iq\left(w^{\prime}+iqw\right)+iq^{\prime}w}
{w^{\prime}+iqw}dx\right]=\int\left(q+\frac{q^{\prime}}{q}
{\rm Re}\left[\frac{1}{y(x)+i}\right]\right)dx.
\end{equation}

The complex-valued admittance $y(x)$ as a function of $\psi$ reads
\begin{equation}\label{eq.11}
y\left[X(\psi)\right]=Y(\psi)=\frac{\dot{W}(\psi)}{\dot{X}(\psi)Q(\psi)W(\psi)}.
\end{equation}

It follows directly from Eq.\ (\ref{eq.9}) that
\begin{equation}\label{eq.10}
\frac{d\psi}{dx}\equiv\frac{1}{\dot{X}(\psi)}=Q(\psi)+
\frac{\dot{G}(\psi)}{\dot{X}(\psi)}\frac{{\rm Re}\left[Y(\psi)\right]}
{\left|Y(\psi)+i\right|^{2}},
\end{equation}
then Eq.\ (1) can be rewritten as a pair of nonlinear differential equations
\begin{eqnarray}
\dot{X}&=&\frac{1}{Q}\left(1-\frac{\dot{G}\: {\rm Re}\: Y}{\left|Y+i\right|^{2}}\right),\label{eq.12_1}\\ 
\dot{Y}&=&\frac{\dot{G}\: {\rm Im}\: Y
\left[i\left(Y^{2}-1\right)-2Y\right]}
{\left|Y+i\right|^{2}}-\left(1+Y^{2}\right),\label{eq.12_2}
\end{eqnarray}
where we use the fact that $\dot{Q}(\psi)=\dot{G}(\psi)Q(\psi)$. The second equation above is obtained if we insert $w^{\prime\prime}(x)=\left[q(x)w(x)h(x)\right]^{\prime}$ into Eq.\ (\ref{eq.1}), i.e.,
\begin{equation}\label{eq.13}
w^{\prime\prime}\left[X(\psi)\right]=
Q^{2}W\left(\frac{\dot{Y}+Y\dot{G}}{Q\dot{X}}+Y^{2}\right)\equiv -Q^{2}W=-q^{2}w.
\end{equation}
This provides us also with the compatibility condition (cf.\ Eq.\ (\ref{eq.1b}))
\begin{equation}\label{eq.13a}
\dot{Y}+Y\dot{G}+\left(1+Y^{2}\right)Q\dot{X}=0
\end{equation}
imposing constraints on choosing the complex admittance $Y(\psi)$.

It is interesting to note that according to Eqs.\ (\ref{eq.12_1}) and (\ref{eq.13a}) we can write that
\begin{equation}\label{eq.14c}
Q\dot{X}=-\frac{\dot{Y}+\dot{G}Y}{1+Y^{2}}=1-\frac{\dot{G}\ {\rm Re}\ Y}{|Y+i|^{2}}
\end{equation}
or equivalently
\begin{equation}\label{eq.14d}
\dot{Y}=\dot{G}\left[
\frac{\left(1+Y^{2}\right)\: {\rm Re}\: Y}{|Y+i|^{2}}-Y\right]
-\left(1+Y^{2}\right).
\end{equation}
This is another way to write Eq.\ (\ref{eq.12_2}). Let us also note that
\begin{equation}\label{eq.12c}
\left|y(x)+i\right|^{2}=\left|y(x)\right|^{2}+2{\rm Im}\left[y(x)\right]+1.
\end{equation}

If the functions $Q(\psi)$ and $Y(\psi)$ are given, then using Eqs.\ (\ref{eq.11}), (\ref{eq.12_1}), (\ref{eq.12_2}), and (\ref{eq.13a}) the following expressions for $X(\psi)$ and the complex-valued wave function $W(\psi)$ can be written:
\begin{eqnarray}
X(\psi)&=&\int\left(1-\frac{\dot{G}\ {\rm Re}\ Y}{|Y+i|^{2}}\right)\frac{d\psi}{Q},\label{eq.14a}\\
W(\psi)&=&w_{0}\exp\left[\int Q\dot{X}Yd\psi\right]=
w_{0}\exp\left[\int \left(1-\frac{\dot{G}\ {\rm Re}\ Y}{|Y+i|^{2}}\right)Yd\psi\right].\quad\label{eq.14b}
\end{eqnarray}

\section{$\mathcal{R},\mathcal{Y}$-variables}

For the sake of convenience, let us denote $Y=\mathcal{R}\exp(i\mathcal{Y})$ and separate real and imaginary parts of Eq.\ (\ref{eq.12_2}) (or equivalently Eq.\ (\ref{eq.14d})), then as a result we obtain the following pair of nonlinear differential equations:
\begin{eqnarray}
\dot{\mathcal{R}}&=&-\dot{G}\: \mathcal{S}
\left(\mathcal{R},\mathcal{Y}\right)\left[
2\mathcal{R}+\left(1+\mathcal{R}^{2}\right)\sin\mathcal{Y}\right]-
\left(1+\mathcal{R}^{2}\right)\cos\mathcal{Y}\nonumber\\
&=&-\dot{G}R+\left(\mathcal{R}^{2}+1\right)\left[
\dot{G}\: \mathcal{C}\left(\mathcal{R},\mathcal{Y}\right)
-1\right]\cos\mathcal{Y},\label{eq.15a}\\
\dot{\mathcal{Y}}&=&\frac{\mathcal{R}^{2}-1}{\mathcal{R}}\left[
\dot{G}\: \mathcal{C}\left(\mathcal{R},\mathcal{Y}\right)
-1\right]\sin\mathcal{Y},\label{eq.15b}
\end{eqnarray}
where
\begin{equation}\label{eq.15b2}
\mathcal{S}\left(\mathcal{R},\mathcal{Y}\right)=
\frac{\mathcal{R}\sin\mathcal{Y}}
{1+\mathcal{R}^{2}+2\mathcal{R}\sin\mathcal{Y}}, \quad \mathcal{C}\left(\mathcal{R},\mathcal{Y}\right)=
\frac{\mathcal{R}\cos\mathcal{Y}}
{1+\mathcal{R}^{2}+2\mathcal{R}\sin\mathcal{Y}}.
\end{equation}

In new variables we have that
\begin{equation}\label{eq.14e}
{\rm Re}\: Y(\psi)=\mathcal{R}(\psi)\cos\mathcal{Y}(\psi),\qquad 
{\rm Im}\: Y(\psi)=\mathcal{R}(\psi)\sin\mathcal{Y}(\psi), 
\end{equation}
and Eq.\ (\ref{eq.12c}) takes the following form:
\begin{equation}\label{eq.12d}
\left|Y(\psi)+i\right|^{2}=1+\mathcal{R}^{2}(\psi)+
2\mathcal{R}(\psi)\sin\mathcal{Y}(\psi).
\end{equation}
Then the functions $\mathcal{S}\left(\mathcal{R},\mathcal{Y}\right)$ and $\mathcal{C}\left(\mathcal{R},\mathcal{Y}\right)$ can be defined as follows:
\begin{equation}\label{eq.14}
\mathcal{S}\left(\mathcal{R},\mathcal{Y}\right)=
\frac{{\rm Im}\: Y(\psi)}{\left|Y(\psi)+i\right|^{2}}, \qquad \mathcal{C}\left(\mathcal{R},\mathcal{Y}\right)=
\frac{{\rm Re}\: Y(\psi)}{\left|Y(\psi)+i\right|^{2}}
\end{equation}
with the compatibility condition
\begin{equation}\label{eq.15}
\mathcal{S}^{2}\left(\mathcal{R},\mathcal{Y}\right)+
\mathcal{C}^{2}\left(\mathcal{R},\mathcal{Y}\right)=
\frac{\left|Y(\psi)\right|^{2}}{\left|Y(\psi)+i\right|^{4}}.
\end{equation}
Then Eq.\ (\ref{eq.12_2}) reads that
\begin{equation}\label{eq.15c}
\left(\dot{\mathcal{R}}+i\mathcal{R}\dot{\mathcal{Y}}\right)e^{i\mathcal{Y}}
=\dot{G}\mathcal{S}\left(\mathcal{R},\mathcal{Y}\right)
\left[i\left(\mathcal{R}^{2}e^{2i\mathcal{Y}}-1\right)-
2\mathcal{R} e^{i\mathcal{Y}}\right]
-\left(1+\mathcal{R}^{2}e^{2i\mathcal{Y}}\right).
\end{equation}
The real and imaginary parts of Eq.\ (\ref{eq.15c}) can be written as follows:
\begin{eqnarray}
\dot{\mathcal{R}}\cos\mathcal{Y}-\mathcal{R}\dot{\mathcal{Y}}\sin\mathcal{Y}&=&
-\dot{G}\mathcal{S}\left(\mathcal{R},\mathcal{Y}\right)\left(1+
\mathcal{R}\sin\mathcal{Y}\right)2\mathcal{R}\cos\mathcal{Y}\nonumber\\
&&-\left(1+\mathcal{R}^{2}\cos 2\mathcal{Y}\right),\label{eq.15d}\\
\dot{\mathcal{R}}\sin\mathcal{Y}+\mathcal{R}\dot{\mathcal{Y}}\cos\mathcal{Y}&=&
-\dot{G}\mathcal{S}\left(\mathcal{R},\mathcal{Y}\right)
\left(1+2\mathcal{R}\sin\mathcal{Y}-\mathcal{R}^{2}\cos 2\mathcal{Y}\right)\nonumber\\
&&-\mathcal{R}^{2}\sin 2\mathcal{Y}.\label{eq.15e}
\end{eqnarray}
Then multiplying Eq.\ (\ref{eq.15d}) by $\cos\mathcal{Y}$ and Eq.\ (\ref{eq.15e}) by $\sin\mathcal{Y}$ and taking their sum we obtain Eq.\ (\ref{eq.15a}) and multiplying Eq.\ (\ref{eq.15d}) by $\sin\mathcal{Y}$ and Eq.\ (\ref{eq.15e}) by $\cos\mathcal{Y}$ and taking their difference we obtain Eq.\ (\ref{eq.15b}). Additionally excluding from Eqs.\ (\ref{eq.15d}) and (\ref{eq.15e}) the term with the derivative $\dot{G}$ we obtain that
\begin{eqnarray}
-\dot{G}\mathcal{S}\left(\mathcal{R},\mathcal{Y}\right)&=&
\frac{\dot{\mathcal{R}}\cos\mathcal{Y}-\mathcal{R}\dot{\mathcal{Y}}\sin\mathcal{Y}
+1+\mathcal{R}^{2}\cos 2\mathcal{Y}}{2\mathcal{R}\cos\mathcal{Y}\left(1+
\mathcal{R}\sin\mathcal{Y}\right)}\nonumber\\
&=&\frac{\dot{\mathcal{R}}\sin\mathcal{Y}+\mathcal{R}\dot{\mathcal{Y}}\cos\mathcal{Y}
+\mathcal{R}^{2}\sin 2\mathcal{Y}}{1+2\mathcal{R}\sin\mathcal{Y}-
\mathcal{R}^{2}\cos 2\mathcal{Y}},\label{eq.15f}
\end{eqnarray}
i.e., equivalently
\begin{equation}\label{eq.15g}
\left[\dot{\mathcal{R}}\cos\mathcal{Y}+\left(1+\mathcal{R}^{2}\right)
+2\mathcal{R}\sin\mathcal{Y}\right]\left(1-\mathcal{R}^{2}\right)=\mathcal{R}\dot{\mathcal{Y}}
\left[2\mathcal{R}+\left(1+\mathcal{R}^{2}\right)\sin\mathcal{Y}\right].
\end{equation}
This is the compatibility condition which is satisfied by $\dot{\mathcal{R}}(\psi)$ and $\dot{\mathcal{Y}}(\psi)$.

Let us note that Eqs.\ (\ref{eq.15a}) and (\ref{eq.15b}) can be equivalently rewritten as follows:
\begin{eqnarray}
\frac{\dot{\mathcal{R}}+\dot{G}\mathcal{R}}{\mathcal{R}^{2}+1}&=&
\left[\dot{G}\: \mathcal{C}\left(\mathcal{R},\mathcal{Y}\right)
-1\right]\cos\mathcal{Y},\label{eq.30a}\\
\frac{\mathcal{R}\dot{\mathcal{Y}}}{\mathcal{R}^{2}-1}&=&
\left[\dot{G}\: \mathcal{C}\left(\mathcal{R},\mathcal{Y}\right)
-1\right]\sin\mathcal{Y}.\label{eq.30b}
\end{eqnarray}

In a general complex case Eqs.\ (\ref{eq.15a}) and (\ref{eq.15b}) can be integrated with respect to $\mathcal{Y}(\psi)$:
\begin{eqnarray}
\mathcal{Y}(\psi)&=&\arcsin\left\{\frac{1+\mathcal{R}^{2}(\psi)}
{\mathcal{R}(\psi)}\: Q(\psi)\exp\left[-2\int 
\frac{\dot{G}(\psi)d\psi}{1+\mathcal{R}^{2}(\psi)}\right]\right\}.
\label{eq.17b}
\end{eqnarray}
Really, let us denote 
\begin{equation}\label{eq.12}
\mathcal{J}\left(\mathcal{R},\mathcal{Y}\right)=
\frac{\mathcal{R}\sin\mathcal{Y}}{\mathcal{R}^{2}+1}.
\end{equation}
Then using Eqs.\ (\ref{eq.15a}) and (\ref{eq.15b}) we can calculate its derivative with respect to $\psi$:
\begin{equation}\label{eq.16}
\dot{\mathcal{J}}\left(\mathcal{R},\mathcal{Y}\right)=
\frac{\left(1-\mathcal{R}^{2}\right)\dot{\mathcal{R}}\sin\mathcal{Y}
+\left(1+\mathcal{R}^{2}\right)\mathcal{R}\dot{\mathcal{Y}}\cos\mathcal{Y}}
{\left(\mathcal{R}^{2}+1\right)^{2}}=
\frac{\mathcal{R}^{2}-1}{\mathcal{R}^{2}+1}\: \dot{G}\mathcal{J}\left(\mathcal{R},\mathcal{Y}\right).
\end{equation}
We see that Eq.\ (\ref{eq.16}) can be easily integrated:
\begin{equation}\label{eq.16a}
\mathcal{J}\left(\psi\right)=\mathcal{J}\left[\mathcal{R}(\psi),
\mathcal{Y}(\psi)\right]=\mathcal{E}(\psi),
\end{equation}
where
\begin{equation}\label{eq.16d}
\mathcal{E}(\psi)=\exp\left[
\int\frac{\mathcal{R}^{2}(\psi)-1}{\mathcal{R}^{2}(\psi)+1}\:
\dot{G}(\psi)d\psi\right]=Q(\psi)\exp\left[-2\int 
\frac{\dot{G}(\psi)d\psi}{1+\mathcal{R}^{2}(\psi)}\right]
\end{equation}
and we used the fact that
\begin{equation}
\int \dot{G}(\psi)d\psi=\int\frac{\dot{Q}(\psi)}{Q(\psi)}d\psi=
\ln Q(\psi).\label{eq.17f1}
\end{equation}
Finally, we obtain that
\begin{equation}\label{eq.17e}
\sin\mathcal{Y}(\psi)=\frac{1+\mathcal{R}^{2}(\psi)}
{\mathcal{R}(\psi)}\: \mathcal{E}(\psi)
\end{equation}
and this leads us directly to Eq.\ (\ref{eq.17b}).

\section{Second-order nonlinear differential equation}

The functions $\mathcal{S}(\psi)=\mathcal{S}\left[\mathcal{R}(\psi),\mathcal{Y}(\psi)\right]$ and $\mathcal{E}(\psi)$ (or $\mathcal{J}(\psi)$ equivalently) are connected by the following relation:
\begin{equation}\label{eq.17f}
\mathcal{S}(\psi)=\frac{\mathcal{E}(\psi)}{1+2\mathcal{E}(\psi)}\qquad \left(=\frac{\mathcal{J}(\psi)}{1+2\mathcal{J}(\psi)}\right).
\end{equation}
We can also calculate the following derivatives with respect to $\psi$:
\begin{eqnarray}
\left[\frac{\mathcal{R}\cos\mathcal{Y}}{\mathcal{R}^{2}+1}\right]^{\cdot}&=&
\frac{\left(1-\mathcal{R}^{2}\right)\dot{\mathcal{R}}\cos\mathcal{Y}
-\left(1+\mathcal{R}^{2}\right)\mathcal{R}\dot{\mathcal{Y}}\sin\mathcal{Y}}
{\left(\mathcal{R}^{2}+1\right)^{2}}\nonumber\\
&=&\frac{\mathcal{R}^{2}-1}{\mathcal{R}^{2}+1}\left\{1+2\dot{G}\: \mathcal{S}\left(\mathcal{R},\mathcal{Y}\right)
\left[\frac{\mathcal{R}\cos\mathcal{Y}}{\mathcal{R}^{2}+1}\right]\right\},
\label{eq.39a}\\
\left[\mathcal{S}\left(\mathcal{R},\mathcal{Y}\right)
\frac{\mathcal{R}\cos\mathcal{Y}}{\mathcal{R}^{2}+1}\right]^{\cdot}&=&
\left[\mathcal{C}\left(\mathcal{R},\mathcal{Y}\right)
\frac{\mathcal{R}\sin\mathcal{Y}}{\mathcal{R}^{2}+1}\right]^{\cdot}
\nonumber\\
&=&\frac{\mathcal{R}^{2}-1}{\mathcal{R}^{2}+1}\left\{\mathcal{S}
\left(\mathcal{R},\mathcal{Y}\right)+
\dot{G}\left[\mathcal{S}\left(\mathcal{R},\mathcal{Y}\right)
\frac{\mathcal{R}\cos\mathcal{Y}}{\mathcal{R}^{2}+1}
\right]\right\}.\qquad\label{eq.39b}
\end{eqnarray}

For the function $\mathcal{C}(\psi)=\mathcal{C}\left[\mathcal{R}(\psi),\mathcal{Y}(\psi)\right]$ we obtain a nice nonlinear second-order differential equation
\begin{equation}\label{eq.38}
\ddot{\mathcal{C}}(\psi)+4\mathcal{C}(\psi)=\frac{\dot{G}(\psi)}{2}
\left[\dot{\mathcal{C}}^{2}(\psi)+4\mathcal{C}^{2}(\psi)-1\right]
\end{equation}
with the eigenfrequency 2 and modulation determined by the variable refraction index $n(x)=n_{0}\exp\left[G\left[\psi(x)\right]\right]$, where $n_{0}=\left(c/\omega\right)q_{0}$.

Really, using Eqs.\ (\ref{eq.15a}) and (\ref{eq.15b}) we can calculate the first and second derivatives of the function $\mathcal{C}(\psi)$ as follows:
\begin{eqnarray}
\dot{\mathcal{C}}&=&
\frac{\left(1-\mathcal{R}^{2}\right)\dot{\mathcal{R}}\cos\mathcal{Y}
-\left[\left(1+\mathcal{R}^{2}\right)\sin\mathcal{Y}+2\mathcal{R}\right]
\mathcal{R}\dot{\mathcal{Y}}
}{\left(1+\mathcal{R}^{2}+2\mathcal{R}\sin\mathcal{Y}\right)^{2}}\nonumber\\
&=&\frac{\mathcal{R}^{2}-1}{1+\mathcal{R}^{2}+2\mathcal{R}\sin\mathcal{Y}}
=1-2\mathcal{S}-\frac{2}{1+\mathcal{R}^{2}+2\mathcal{R}\sin\mathcal{Y}}
\label{eq.16b}
\end{eqnarray}
and
\begin{eqnarray}
\ddot{\mathcal{C}}&=&-2\dot{\mathcal{S}}
-2\left[\frac{1}{1+\mathcal{R}^{2}+2\mathcal{R}\sin\mathcal{Y}}\right]^{\cdot}.
\label{eq.16e}
\end{eqnarray}
Calculating the auxiliary derivatives
\begin{eqnarray}
\dot{\mathcal{S}}&=&
\frac{\left(1-\mathcal{R}^{2}\right)\dot{\mathcal{R}}\sin\mathcal{Y}
+\left(1+\mathcal{R}^{2}\right)\mathcal{R}\dot{\mathcal{Y}}\cos\mathcal{Y}}
{\left(1+\mathcal{R}^{2}+2\mathcal{R}\sin\mathcal{Y}\right)^{2}}\nonumber\\
&=&\dot{G}\mathcal{S}-2\dot{G}\mathcal{S}^{2}-2\dot{G}\mathcal{S}
\left[\frac{1}{1+\mathcal{R}^{2}+2\mathcal{R}\sin\mathcal{Y}}\right]
\label{eq.16f}
\end{eqnarray}
and
\begin{equation}\label{eq.16g}
\left[\frac{1}{1+\mathcal{R}^{2}+2\mathcal{R}\sin\mathcal{Y}}\right]^{\cdot}=
2\mathcal{C}+2\dot{G}\mathcal{S}^{2}+2\dot{G}\mathcal{S}
\left[\frac{1}{1+\mathcal{R}^{2}+2\mathcal{R}\sin\mathcal{Y}}\right],
\end{equation}
we see that their second and third terms are eliminating each other and then
\begin{equation}\label{eq.16h}
\ddot{\mathcal{C}}=-4\mathcal{C}-2\dot{G}\mathcal{S}.
\end{equation}
Let us also note that
\begin{eqnarray}
\dot{\mathcal{C}}^{2}+4\mathcal{C}^{2}&=&\frac{\left(
\mathcal{R}^{2}-1\right)^{2}+4\mathcal{R}^{2}\cos^{2}\mathcal{Y}}{\left(
\mathcal{R}^{2}+1\right)^{2}\left(1+2\mathcal{J}\right)^{2}}=\frac{\left(
\mathcal{R}^{2}+1\right)^{2}-4\mathcal{R}^{2}\sin^{2}\mathcal{Y}}{\left(
\mathcal{R}^{2}+1\right)^{2}\left(1+2\mathcal{J}\right)^{2}}\nonumber\\
&=&\frac{1-4\mathcal{J}^{2}}{\left(1+2\mathcal{J}\right)^{2}}=
\frac{1-2\mathcal{J}}{1+2\mathcal{J}}=1-4\mathcal{S}.\label{eq.16k}
\end{eqnarray}
Finally, excluding $\mathcal{S}$ from Eqs.\ (\ref{eq.16h}) and (\ref{eq.16k}) we obtain Eq.\ (\ref{eq.38}). 

There are two ways of constructing sought parametric solutions: 
\begin{itemize}
\item[$(i)$] to define $G(\psi)$ and then solve Eq.\ (\ref{eq.38}) with respect to $\mathcal{C}(\psi)$ or 

\item[$(ii)$] to define $\mathcal{C}(\psi)$ and then find $G(\psi)$ by integration:
\begin{equation}\label{eq.19}
G(\psi)=2\int\frac{\ddot{\mathcal{C}}(\psi)+4\mathcal{C}(\psi)}{
\dot{\mathcal{C}}^{2}(\psi)+4\mathcal{C}^{2}(\psi)-1}\: d\psi.
\end{equation}
\end{itemize}

Let us note that for any function $\dot{G}(\psi)$ there are two particular solutions of Eq.\ (\ref{eq.38}), namely,
\begin{equation}\label{eq.19a}
\mathcal{C}_{1}(\psi)=\alpha\sin\beta\psi,\qquad \mathcal{C}_{2}(\psi)=\alpha\cos\beta\psi,
\end{equation}
where
\begin{equation}\label{eq.19j}
\alpha=\pm\frac{1}{\beta},\qquad \beta=\pm 2.
\end{equation}
Really, it is easy to check that the first particular solution $\mathcal{C}_{1}(\psi)$ is the solution of Eq.\ (\ref{eq.38}) by direct calculations of the following terms:
\begin{eqnarray}
\ddot{\mathcal{C}}_{1}(\psi)+4\mathcal{C}_{1}(\psi)&=&
\left(4-\beta^{2}\right)\alpha\sin\beta\psi,\label{eq.19b}\\
\dot{\mathcal{C}}_{1}^{2}(\psi)+4\mathcal{C}_{1}^{2}(\psi)-1&=&
\left(4-\beta^{2}\right)\alpha^{2}\sin^{2}\beta\psi+
\left(\alpha^{2}\beta^{2}-1\right).\label{eq.19c}
\end{eqnarray}
Therefore, if we suppose that $\alpha$ and $\beta$ fulfil the following conditions:
\begin{equation}\label{eq.19d}
\alpha^{2}\beta^{2}-1=0,\qquad 4-\beta^{2}=0,
\end{equation}
then for any function $\dot{G}(\psi)$ the left- and right-hand sides of Eq.\ (\ref{eq.38}) are equal to zero separately. The same is true for the second particular solution $\mathcal{C}_{2}(\psi)$.

If we differentiate Eq.\ (\ref{eq.19e}) with respect to $\psi$, then we obtain that
\begin{equation}\label{eq.19f}
2\dot{\mathcal{C}}(\psi)\left(\ddot{\mathcal{C}}(\psi)+
4\mathcal{C}(\psi)\right)=0.
\end{equation}
If $\dot{\mathcal{C}}(\psi)\equiv 0$, then the function $\mathcal{C}(\psi)$ is constant and from Eq.\ (\ref{eq.19}) we obtain
\begin{equation}\label{eq.19g}
G(\psi)=\frac{8C}{4C^{2}-1}\left(\psi-\psi_{0}\right).
\end{equation}
If we take a non-constant ($\dot{\mathcal{C}}(\psi)\neq 0$) solution 
\begin{equation}\label{eq.19h}
\mathcal{C}(\psi)=\pm\frac{1}{2}\sin 2\left(\psi-\psi_{0}\right) 
\end{equation}
of the following differential equation:
\begin{equation}\label{eq.19e}
\dot{\mathcal{C}}^{2}(\psi)+4\mathcal{C}^{2}(\psi)-1=0,
\end{equation}
then Eq.\ (\ref{eq.38}) is fulfilled identically for any function $\dot{G}(\psi)$.

The variables $\mathcal{R}$ and $\mathcal{Y}$ can be expressed through $\mathcal{C}$ and its first derivative $\dot{\mathcal{C}}$ as follows:
\begin{equation}\label{eq.17}
{\rm ctg}\: \mathcal{Y}=\frac{\mathcal{C}}{\mathcal{S}}=\frac{4\mathcal{C}}{1
-4\mathcal{C}^{2}-\dot{\mathcal{C}}^{2}},\qquad \mathcal{R}^{2}=\frac{4\mathcal{C}^{2}+\left(1+\dot{\mathcal{C}}\right)^{2}}{
4\mathcal{C}^{2}+\left(1-\dot{\mathcal{C}}\right)^{2}}.
\end{equation}
Really, using Eqs.\ (\ref{eq.15b2}), (\ref{eq.12}), (\ref{eq.17f}), and (\ref{eq.16k}) we can calculate that 
\begin{eqnarray}
\frac{\mathcal{R}\sin\mathcal{Y}}{1+\mathcal{R}^{2}}&=&\mathcal{J}
=\frac{\mathcal{S}}{1-2\mathcal{S}}=\frac{1}{2}\left(
\frac{1-4\mathcal{C}^{2}-\dot{\mathcal{C}}^{2}}{1
+4\mathcal{C}^{2}+\dot{\mathcal{C}}^{2}}\right),\label{eq.17a1}\\ 
\frac{\mathcal{R}\cos\mathcal{Y}}{1+\mathcal{R}^{2}}&=&
\mathcal{C}\left(1+2\mathcal{J}\right)=\frac{\mathcal{C}}{1-2\mathcal{S}}=
\frac{2\mathcal{C}}{1+4\mathcal{C}^{2}+\dot{\mathcal{C}}^{2}},\label{eq.17a2}
\end{eqnarray}
from where the first of Eqs.\ (\ref{eq.17}) follows directly. 

Additionally, using Eq.\ (\ref{eq.16b}) we obtain that
\begin{equation}\label{eq.17a}
\frac{\mathcal{R}^{2}-1}{\mathcal{R}^{2}+1}=
\dot{\mathcal{C}}\left(1+2\mathcal{J}\right)=
\frac{\dot{\mathcal{C}}}{1-2\mathcal{S}}=
\frac{2\dot{\mathcal{C}}}{1+4\mathcal{C}^{2}+\dot{\mathcal{C}}^{2}}.
\end{equation}
Solving Eq.\ (\ref{eq.17a}) with respect to $\mathcal{R}^{2}$ provides us with the second of Eqs.\ (\ref{eq.17}).

Therefore, the complex admittance $Y=\mathcal{R}\exp(i\mathcal{Y})$ can be expressed through $\mathcal{C}$ and its first derivative $\dot{\mathcal{C}}$ as follows:
\begin{equation}\label{eq.18a}
Y=\left(1+\mathcal{R}^{2}\right)
\left[\frac{\mathcal{R}\cos\mathcal{Y}}{1+\mathcal{R}^{2}}+
i\: \frac{\mathcal{R}\sin\mathcal{Y}}{1+\mathcal{R}^{2}}\right]=
\frac{4\mathcal{C}+i\left(1-4\mathcal{C}^{2}-\dot{\mathcal{C}}^{2}\right)}{
4\mathcal{C}^{2}+\left(1-\dot{\mathcal{C}}\right)^{2}}.
\end{equation}

If the functions $Q(\psi)$ and $\mathcal{C}(\psi)$ (or only one of them) are given, then the following expressions for $X(\psi)$ and the complex-valued wave function $W(\psi)$ can be written:
\begin{eqnarray}
X(\psi)&=&\int\left(1-\dot{G}(\psi) \mathcal{C}(\psi)\right)\frac{d\psi}{Q(\psi)},\label{eq.18b}\\
W(\psi)&=&w_{0}\exp\left[\int\left(1-\dot{G}(\psi) \mathcal{C}(\psi)\right)Y(\psi)d\psi\right]\nonumber\\
&=&w_{0}\exp\left[\int\frac{\left(1-\dot{G}\mathcal{C}\right)
\left[4\mathcal{C}+i\left(1-4\mathcal{C}^{2}-\dot{\mathcal{C}}^{2}
\right)\right]}{4\mathcal{C}^{2}+\left(1-\dot{\mathcal{C}}\right)^{2}}\:
d\psi\right].\label{eq.18c}
\end{eqnarray}

For the partial solutions of Eq.\ (\ref{eq.38}), which are given by Eq.\ (\ref{eq.19h}), the admittance $Y(\psi)$ is purely real, i.e.,
\begin{eqnarray}\label{eq.41}
Y(\psi)=\left\{
\begin{array}{lr}
``+": &  {\rm ctg}\left(\psi-\psi_{0}\right),\\
``-": & -{\rm tg}\left(\psi-\psi_{0}\right),
\end{array}
\right.
\end{eqnarray}
therefore, for any given function $\dot{G}(\psi)$ (equivalently $Q(\psi)$) we obtain the following expressions for $X(\psi)$ and the complex-valued wave function $W(\psi)$:
\begin{eqnarray}
X(\psi)&=&\int\left[1\mp\dot{G}(\psi)\sin\left(\psi-\psi_{0}\right)
\cos\left(\psi-\psi_{0}\right)\right]\frac{d\psi}{Q(\psi)},\label{eq.42a}\\
W(\psi)&=&w_{0}\sqrt{1-Z^{2}(\psi)}
\exp\left[-\int\dot{G}(\psi)Z^{2}(\psi)d\psi\right],\label{eq.42b}
\end{eqnarray}
where
\begin{eqnarray}\label{eq.42}
Z(\psi)=\left\{
\begin{array}{lr}
``+": & \cos\left(\psi-\psi_{0}\right),\\
``-": & \sin\left(\psi-\psi_{0}\right).
\end{array}
\right.
\end{eqnarray}
Really, to obtain Eq.\ (\ref{eq.42b}) we need to calculate the following integral:
\begin{equation}\label{eq.42c}
\int\left(1-\dot{G}(\psi)\mathcal{C}(\psi)\right)Y(\psi)d\psi,
\end{equation}
where $\mathcal{C}(\psi)$ is given by Eq.\ (\ref{eq.19h}), and for the complex admittance we obtain Eq.\ (\ref{eq.41}) by direct calculation. Then Eq.\ (\ref{eq.42c}) can be rewritten as follows:
\begin{equation}\label{eq.42d}
\int Y(\psi)d\psi - \int\dot{G}(\psi)\mathcal{C}(\psi)Y(\psi)d\psi=
\ln\sqrt{1-Z^{2}(\psi)}-\int\dot{G}(\psi)Z^{2}(\psi)d\psi,
\end{equation}
where
\begin{eqnarray}\label{eq.42e}
\sqrt{1-Z^{2}(\psi)}=\left\{
\begin{array}{lr}
``+": & \sin\left(\psi-\psi_{0}\right),\\
``-": & \cos\left(\psi-\psi_{0}\right).
\end{array}
\right.
\end{eqnarray}
Finally, we see that Eq.\ (\ref{eq.42b}) directly follows from Eqs.\ (\ref{eq.42d}) and (\ref{eq.42e}).

For a constant function $\mathcal{C}(\psi)$ and $G(\psi)$ given by Eq.\ (\ref{eq.19g}), we obtain the following expressions for $Q(\psi)$, the complex-valued admittance $Y(\psi)$, $X(\psi)$, and the complex-valued wave function $W(\psi)$:
\begin{eqnarray}
Q(\psi)&=&q_{0}\exp\left[\frac{8\mathcal{C}}{
4\mathcal{C}^{2}-1}\left(\psi-\psi_{0}\right)\right],\label{eq.43_1}\\
Y(\psi)&=&\frac{4\mathcal{C}+i\left(1-4\mathcal{C}^{2}\right)}
{1+4\mathcal{C}^{2}},\label{eq.43}\\
X(\psi)&=&\frac{1+4\mathcal{C}^{2}}{8\mathcal{C}}\frac{1}{Q(\psi)},
\label{eq.43a}\\
W(\psi)&=&w_{0}\exp\left[\left\{\frac{4\mathcal{C}}{1-4\mathcal{C}^{2}}+
i\right\}\left(\psi-\psi_{0}\right)\right].\label{eq.43b}
\end{eqnarray} 
Eqs.\ (\ref{eq.43a}) and (\ref{eq.43b}) directly follows from the fact that the expression
\begin{equation}\label{eq.43c}
1-\dot{G}(\psi)\mathcal{C}(\psi)=1-\frac{8\mathcal{C}^{2}}{
4\mathcal{C}^{2}-1}=\frac{1+4\mathcal{C}^{2}}{1-4\mathcal{C}^{2}}
\end{equation}
and complex-valued admittance $Y(\psi)$, given by Eq.\ (\ref{eq.43}), are constants.

\section{Special solutions}

Though it is hardly possible to find the exact solution of Eq.\ (\ref{eq.38}) in a general case, the above analysis clarifies the nature of quasi-periodic Bloch waves in the transmission band and allows one to construct a wide class of special analytical solutions. A continual set of integrable wave equations can be obtained if we choose
\begin{equation}\label{eq.29}
\dot{G}\left[\psi(\mathcal{C})\right]=\frac{d\ln M(\mathcal{C})}{d\mathcal{C}}=\frac{1}{M(\mathcal{C})}
\frac{dM(\mathcal{C})}{d\mathcal{C}}, 
\end{equation}
where $M(\mathcal{C})$ is an arbitrary real-valued function. In this case Eq.\ (\ref{eq.38}) has an energy integral
\begin{equation}\label{eq.22}
\dot{\mathcal{C}}^{2}=1-4\mathcal{C}^{2}+M(\mathcal{C})
\end{equation}
and a periodic solution $\mathcal{C}(\psi)=\mathcal{C}(\psi+\tau)$ given by the following expressions:
\begin{equation}\label{eq.23}
\psi=\pm\int\frac{d\mathcal{C}}{
\sqrt{1-4\mathcal{C}^{2}+M(\mathcal{C})}},\qquad \tau=2
\int^{\mathcal{C}_{+}}_{\mathcal{C}_{-}}
\frac{d\mathcal{C}}{\sqrt{1-4\mathcal{C}^{2}+M(\mathcal{C})}},
\end{equation}
where the turning points $\mathcal{C}_{\pm}$ are the roots of the radical. 

Let us notice that using Eq.\ (\ref{eq.29}) we can calculate the complete derivative of $M\left[\mathcal{C}(\psi)\right]$ with respect to $\psi$ as follows:
\begin{equation}\label{eq.29a}
\frac{\dot{M}(\mathcal{C})}{M(\mathcal{C})}=
\frac{1}{M(\mathcal{C})}\frac{dM(\mathcal{C})}{d\mathcal{C}}
\frac{d\mathcal{C}}{d\psi}=\dot{G}\dot{\mathcal{C}}.
\end{equation}
Then rewriting Eq.\ (\ref{eq.38}) in the form
\begin{equation}\label{eq.29b}
\dot{G}\dot{\mathcal{C}}=\frac{2\dot{\mathcal{C}}\left(\ddot{\mathcal{C}}+
4\mathcal{C}\right)}{\dot{\mathcal{C}}^{2}+4\mathcal{C}^{2}-1}
=\frac{d}{d\psi}\ln\left[\dot{\mathcal{C}}^{2}+4\mathcal{C}^{2}-1\right]
\equiv \frac{d}{d\psi}\ln M(\mathcal{C})
\end{equation}
and integrating Eq.\ (\ref{eq.29b}) we obtain that
\begin{equation}\label{eq.29e}
\frac{\dot{\mathcal{C}}^{2}+4\mathcal{C}^{2}-1}{M(\mathcal{C})}={\rm const}.
\end{equation}
Due to the arbitrariness of the very function $M(\mathcal{C})$, we can leave out the constant on the right-hand side of Eq.\ (\ref{eq.29e}). This leads us directly to Eq.\ (\ref{eq.22}).

Eq.\ (\ref{eq.38}) can be written in the following form:
\begin{equation}\label{eq.44}
\ddot{\mathcal{C}}=f\left(\dot{\mathcal{C}},\mathcal{C},\psi\right),
\end{equation}
where the direct dependence on $\psi$ is realized only through the function $\dot{G}(\psi)$. Let us suppose that in some way we have rewritten it as a function of $\mathcal{C}$, i.e., $\dot{G}(\mathcal{C})=\dot{G}\left[\psi(\mathcal{C})\right]$. Then in Eq.\ (\ref{eq.44}) the direct dependence on $\psi$ is missing, and therefore, we can take $\mathcal{C}$ as an independent variable. Then we obtain that $\dot{\mathcal{C}}=y(\mathcal{C})$, $\ddot{\mathcal{C}}=y(\mathcal{C})y^{\prime}(\mathcal{C})$, and Eq.\ (\ref{eq.44}) reads
\begin{equation}\label{eq.44a}
2y(\mathcal{C})y^{\prime}(\mathcal{C})-\dot{G}(\mathcal{C})y^{2}(\mathcal{C})=
\dot{G}(\mathcal{C})\left(4\mathcal{C}^{2}-1\right)-8\mathcal{C},
\end{equation}
where $\prime$ denotes the derivative with respect to $\mathcal{C}$. Let us note that
\begin{equation}\label{eq.44b}
\left(\frac{y^{2}}{M(\mathcal{C})}\right)^{\prime}=
\frac{1}{M(\mathcal{C})}\left[2yy^{\prime}-
\frac{M^{\prime}(\mathcal{C})}{M(\mathcal{C})}y^{2}\right],
\end{equation}
and we see that to integrate Eq.\ (\ref{eq.44a}) it is enough to suppose that the connection between the functions $\dot{G}(\mathcal{C})$ and $M(\mathcal{C})$ is given by Eq.\ (\ref{eq.29}). Then we obtain the following first-order differential equation:
\begin{equation}\label{eq.44c}
\dot{\mathcal{C}}^{2}=y^{2}=M(\mathcal{C})\left[
\int\frac{4\mathcal{C}^{2}-1}{M^{2}(\mathcal{C})}\: dM(\mathcal{C})
-\int \frac{8\mathcal{C}}{M(\mathcal{C})}\: d\mathcal{C}\right].
\end{equation}
If we compare Eq.\ (\ref{eq.44c}) with Eq.\ (\ref{eq.22}), we obtain the compatibility condition
\begin{equation}\label{eq.44d}
\frac{4\mathcal{C}^{2}-1}{M(\mathcal{C})}+
\int\frac{4\mathcal{C}^{2}-1}{M^{2}(\mathcal{C})}\: dM(\mathcal{C})
=1+\int \frac{8\mathcal{C}}{M(\mathcal{C})}\: d\mathcal{C}.
\end{equation}

If we suppose that the function $M(\mathcal{C})$ is constant, i.e.,
\begin{equation}\label{eq.41a}
M(\mathcal{C})=c,\qquad c>-1,\qquad c\neq 0, 
\end{equation}
then $\dot{G}=0$ and we obtain that 
\begin{equation}\label{eq.39}
\psi(\mathcal{C})=\pm\int\frac{d\mathcal{C}}{
\sqrt{1+c-4\mathcal{C}^{2}}}=\pm\frac{1}{2}\int\frac{dy}{
\sqrt{1-y^{2}}},\qquad y=\frac{2\mathcal{C}}{\sqrt{1+c}}.
\end{equation}
The integral in Eq.\ (\ref{eq.39}) gives us the ${\rm arcsin}$ function, therefore, after the inversion we finally obtain that the auxiliary function $\mathcal{C}(\psi)$ fulfilling Eq.\ (\ref{eq.38}) has the following form:
\begin{equation}\label{eq.40}
\mathcal{C}(\psi)=\pm\frac{\sqrt{1+c}}{2}\sin 2\left(\psi-\psi_{0}\right),
\qquad \psi_{0}=\psi(0).
\end{equation}

If we suppose that
\begin{equation}\label{eq.41b}
M(\mathcal{C})=c+8e\mathcal{C},\qquad c>-1-4e^{2},\qquad c\neq 0,
\end{equation}
where $c$ and $e$ are constants, then 
\begin{equation}\label{eq.39e}
\psi(\mathcal{C})=\pm\int\frac{d\mathcal{C}}{
\sqrt{1+c+8e\mathcal{C}-4\mathcal{C}^{2}}}=\pm\frac{1}{2}\int\frac{dy}{
\sqrt{1-y^{2}}},\ \ y=\frac{2\left(\mathcal{C}-e\right)}{\sqrt{1+c+4e^{2}}}.\
\end{equation}
After the integration and inversion of Eq.\ (\ref{eq.39e}) with respect to the auxiliary function $\mathcal{C}(\psi)$, we obtain that
\begin{equation}\label{eq.40a}
\mathcal{C}(\psi)=e\pm\frac{\sqrt{1+c+4e^{2}}}{2}\sin 2\left(\psi-\psi_{0}\right).
\end{equation}

If we suppose that
\begin{equation}\label{eq.41c}
M(\mathcal{C})=c+8e\mathcal{C}-d^{2}\mathcal{C}^{2},\qquad c>-1-\frac{16e^{2}}{d^{2}+4},\qquad c\neq 0,
\end{equation}
where $c$, $e$, and $d$ are constants, then 
\begin{eqnarray}
\psi(\mathcal{C})&=&\pm\int\frac{d\mathcal{C}}{
\sqrt{1+c+8e\mathcal{C}-\left(d^{2}+4\right)\mathcal{C}^{2}}}=
\pm\frac{1}{2}\int\frac{dy}{\sqrt{1-y^{2}}},\label{eq.39d}\\
y&=&\frac{\left(d^{2}+4\right)\mathcal{C}-4e}{\sqrt{
(1+c)\left(d^{2}+4\right)+16e^{2}}}.\label{eq.39f}
\end{eqnarray}
After the integration and inversion of Eq.\ (\ref{eq.39d}) we obtain that
\begin{equation}\label{eq.40b}
\mathcal{C}(\psi)=\frac{1}{d^{2}+4}\left\{
4e\pm\sqrt{(1+c)\left(d^{2}+4\right)+16e^{2}}
\sin\left[\sqrt{d^{2}+4}\left(\psi-\psi_{0}\right)\right]\right\}.
\end{equation}

If we suppose that
\begin{equation}\label{eq.41d}
M(\mathcal{C})=c+8e\mathcal{C}+\left(k^{2}+4\right)\mathcal{C}^{2},\qquad c>-1+\frac{16e^{2}}{k^{2}},\quad k>0,\quad c\neq 0,
\end{equation}
where $c$, $e$, and $k$ are constants, then 
\begin{eqnarray}
\psi(\mathcal{C})&=&\pm\int\frac{d\mathcal{C}}{
\sqrt{1+c+8e\mathcal{C}+k^{2}\mathcal{C}^{2}}}=
\pm\frac{1}{2}\int\frac{dy}{\sqrt{1+y^{2}}},\label{eq.39g}\\
y&=&\frac{k^{2}\mathcal{C}+4e}{\sqrt{
(1+c)k^{2}-16e^{2}}}.\label{eq.39h}
\end{eqnarray}
The integral in Eq.\ (\ref{eq.39g}) gives us the hyperbolic ${\rm arcsh}$ function, therefore, after the inversion we finally obtain that
\begin{equation}\label{eq.40c}
\mathcal{C}(\psi)=\frac{1}{k^{2}}\left\{-
4e\pm\sqrt{(1+c)k^{2}-16e^{2}}\:
{\rm sh}\: k\left(\psi-\psi_{0}\right)\right\}.
\end{equation}

Let us also consider an instructive example of modulated waves in a periodic dielectric medium, determined by the following potential:
\begin{equation}\label{eq.21}
M(\mathcal{C})=\left(4a^2-1\right)+b^2\mathcal{C}^4,\qquad ab< 1,\qquad 4a^{2}\neq 1,
\end{equation}
where $a$ and $b$ are some constants ($a,b>0$). Then the auxiliary function $\mathcal{C}(\psi)$ is expressed through the Jacobi elliptic functions of the first kind \cite{5}:
\begin{equation}\label{eq.27}
\mathcal{C}(\psi)=\pm a\sqrt{1+p^{2}}\ {\rm sn}\left[\frac{2\left(\psi-\psi_{0}\right)}{\sqrt{1+p^{2}}},p\right],
\qquad p=\frac{1-\sqrt{1-a^{2}b^{2}}}{ab}.
\end{equation}
Really, Inserting Eq.\ (\ref{eq.21}) into the first of Eqs.\ (\ref{eq.23}) we obtain that
\begin{equation}\label{eq.23a}
\psi(\mathcal{C})=\psi_{0}\pm\int^{\mathcal{C}}_{0}\frac{d\mathcal{C}}{
\sqrt{4a^{2}-4\mathcal{C}^{2}+b^{2}\mathcal{C}^{4}}}=\psi_{0}\pm\frac{1}{b}
\int^{\mathcal{C}}_{0}\frac{d\mathcal{C}}{
\sqrt{\left(\mathcal{C}^{2}_{+}-\mathcal{C}^{2}\right)
\left(\mathcal{C}^{2}_{-}-\mathcal{C}^{2}\right)}},
\end{equation}
where the roots of the radical are given as follows:
\begin{equation}\label{eq.23b}
\mathcal{C}^{2}_{\pm}=\frac{2}{b^{2}}\left(1\pm\sqrt{1-a^{2}b^{2}}\right).
\end{equation}
If we take that $\mathcal{C}_{+}>\mathcal{C}_{-}>\mathcal{C}>0$ (the roots $\mathcal{C}_{\pm}$ are real for $ab\leq 1$; additionally the condition $\mathcal{C}_{+}>\mathcal{C}_{-}$ imposes $ab\neq 1$), then the integral in Eq.\ (\ref{eq.23a}) can be expressed through the inverse Jacobi elliptic functions of the first kind:
\begin{equation}\label{eq.23c}
\psi(\mathcal{C})=\psi_{0}\pm\frac{1}{b\mathcal{C}_{+}}{\rm sn}^{-1}\left(
\frac{\mathcal{C}}{\mathcal{C}_{-}},
\frac{\mathcal{C}_{-}}{\mathcal{C}_{+}}\right).
\end{equation}
Let us denote 
\begin{equation}\label{eq.23d}
\frac{\mathcal{C}_{-}}{\mathcal{C}_{+}}=p=\frac{\sqrt{1-\sqrt{1-a^{2}b^{2}}}
}{\sqrt{1+\sqrt{1-a^{2}b^{2}}}}=\frac{1-\sqrt{1-a^{2}b^{2}}}{ab}.
\end{equation}
Then inverting Eq.\ (\ref{eq.23c}) and taking into account that ${\rm sn}\left(-x\right)=-{\rm sn}\: x$ and that
\begin{equation}\label{eq.23e}
\sqrt{1+p^{2}}=\frac{\sqrt{2}}{ab}\sqrt{1-\sqrt{1-a^{2}b^{2}}}=
\frac{\mathcal{C}_{-}}{a},\qquad \mathcal{C}_{+}\mathcal{C}_{-}=\frac{2a}{b},
\end{equation}
we finally obtain Eq.\ (\ref{eq.27}).

For any given function $M\left[\mathcal{C}(\psi)\right]$ the complex-valued wave function $W(\psi)$ can be rewritten as follows:
\begin{equation}\label{eq.18}
W=w_{0}\exp\left[\pm\int\frac{4\mathcal{C}-iM}{
4\mathcal{C}^{2}+\left(1\mp\sqrt{1-4\mathcal{C}^{2}+M}\right)^{2}}\: 
\frac{\left(M-\mathcal{C}M^{\prime}\right)d\mathcal{C}}{
M\sqrt{1-4\mathcal{C}^{2}+M}}\right].
\end{equation}
Really, on the period we have to take the sign "$+$" while integrating from $\mathcal{C}_{-}$ to $\mathcal{C}_{+}$ and "$-$" while integrating back from $\mathcal{C}_{+}$ to $\mathcal{C}_{-}$ in the integral in Eq.\ (\ref{eq.18}):
\begin{eqnarray}
&&\int^{\mathcal{C}_{+}}_{\mathcal{C}_{-}}
\frac{4\mathcal{C}-iM}{
4\mathcal{C}^{2}+\left(1-\sqrt{1-4\mathcal{C}^{2}+M}\right)^{2}}\: 
\frac{\left(M-\mathcal{C}M^{\prime}\right)d\mathcal{C}}{
M\sqrt{1-4\mathcal{C}^{2}+M}}\nonumber\\
&-&\int^{\mathcal{C}_{-}}_{\mathcal{C}_{+}}
\frac{4\mathcal{C}-iM}{
4\mathcal{C}^{2}+\left(1+\sqrt{1-4\mathcal{C}^{2}+M}\right)^{2}}\: 
\frac{\left(M-\mathcal{C}M^{\prime}\right)d\mathcal{C}}{
M\sqrt{1-4\mathcal{C}^{2}+M}}.\label{eq.25a}
\end{eqnarray}
Then we can rewrite Eq.\ (\ref{eq.25a}) as follows:
\begin{eqnarray}
&&\int^{\mathcal{C}_{+}}_{\mathcal{C}_{-}}
\frac{4\mathcal{C}-iM}{2+M-2\sqrt{1-4\mathcal{C}^{2}+M}}
\frac{\left(M-\mathcal{C}M^{\prime}\right)d\mathcal{C}}{
M\sqrt{1-4\mathcal{C}^{2}+M}}\nonumber\\
&+&\int^{\mathcal{C}_{+}}_{\mathcal{C}_{-}}
\frac{4\mathcal{C}-iM}{2+M+2\sqrt{1-4\mathcal{C}^{2}+M}} 
\frac{\left(M-\mathcal{C}M^{\prime}\right)d\mathcal{C}}{
M\sqrt{1-4\mathcal{C}^{2}+M}}\nonumber\\
&=&2\int^{\mathcal{C}_{+}}_{\mathcal{C}_{-}}
\frac{\left(2+M\right)}{
\left(2+M\right)^{2}-4+16\mathcal{C}^{2}-4M}
\left\{\frac{4\mathcal{C}}{M}-i\right\}
\frac{\left(M-\mathcal{C}M^{\prime}\right)d\mathcal{C}}{
\sqrt{1-4\mathcal{C}^{2}+M}}.\qquad\label{eq.25b}
\end{eqnarray}
And this leads us directly to Eq.\ (\ref{eq.25}).

In particular, for the complex increment we obtain that
\begin{eqnarray}
\chi+i\eta&=&\ln\left[\frac{W(\psi+\tau)}{W(\psi)}\right]\nonumber\\
&=&2\int^{\mathcal{C}_{+}}_{\mathcal{C}_{-}}
\frac{2+M}{M^{2}+16\mathcal{C}^{2}}\left\{\frac{4\mathcal{C}}{M}-i\right\} \frac{\left(M-\mathcal{C}M^{\prime}\right)d\mathcal{C}}{
\sqrt{1-4\mathcal{C}^{2}+M}}.\label{eq.25}
\end{eqnarray}

Let us take an arbitrary even function $M(\mathcal{C})$, then the function
\begin{equation}\label{eq.24}
G(\psi)=\int\dot{G}(\psi)d\psi=\pm\int\frac{dM(\mathcal{C})}{M(\mathcal{C})
\sqrt{1-4\mathcal{C}^{2}+M(\mathcal{C})}}
\end{equation}
will be periodic. Moreover, for any even function $M(\mathcal{C})$ we have $\chi=0$, which means that $|W(\psi)|$ is periodic, while the phase advance per period $\tau$, i.e.,
\begin{equation}\label{eq.26}
\eta=4\int^{\mathcal{C}_{+}}_{0}\frac{2+M}{M^{2}+16\mathcal{C}^{2}} \frac{\left(\mathcal{C}M^{\prime}-M\right)d\mathcal{C}}{
\sqrt{1-4\mathcal{C}^{2}+M}},
\end{equation}
determines the modulation period $T=\left(2\pi/\eta\right)\tau$ of the quasi-periodic solution $W(\psi)$ predicted by the Floquet's theory.

For any even function $M(\mathcal{C})$ the real part $\chi$ of the complex increment given by Eq.\ (\ref{eq.25}), i.e.,
\begin{equation}\label{eq.45}
\chi=2\int^{\mathcal{C}_{+}}_{\mathcal{C}_{-}}
\frac{2+M}{M^{2}+16\mathcal{C}^{2}}
\left\{1-\frac{\mathcal{C}M^{\prime}}{M}\right\} 
\frac{4\mathcal{C}d\mathcal{C}}{\sqrt{1-4\mathcal{C}^{2}+M}},
\end{equation}
is an integral from $\mathcal{C}_{-}$ to $\mathcal{C}_{+}$ of the expression which is odd. Therefore, if we take the symmetric range of integration, then the integral of the odd expression vanishes and $\chi=0$. The purely imaginary part $\eta$ of the complex increment is an integral of the expression which is even, hence, taking the symmetric range of integration gives us exactly Eq.\ (\ref{eq.26}).

%\section*{Conclusions}

\section*{Acknowledgements}

This paper partially contains results obtained within the framework of the research project N N501 049 540 financed from the Scientific Research Support Fund in 2011-2014. The second author (VK) is greatly indebted to the Polish Ministry of Science and Higher Education for this financial support.

\end{document}